\documentclass[12pt,a4paper]{article}
\usepackage{graphicx}
\usepackage{epsf}
\usepackage{times}
\title{\bf The most luminous stars \\ 
in the Galaxy and the Magellanic Clouds}
\author{Wolf-Rainer Hamann, 
Andreas Barniske,
Adriane Liermann,\\
Lidia M. Oskinova, 
Diana Pasemann,
Ute R\"uhling\\
\vspace{1cm}\\
Universit\"at Potsdam, Germany
}
\date{\mbox{}}
\begin{document}
\maketitle
\pagestyle{empty}
%
%
\def\bull{\vrule height .9ex width .8ex depth -.1ex}
\makeatletter
\def\ps@plain{\let\@mkboth\gobbletwo
\def\@oddhead{}\def\@oddfoot{\hfil\tiny\bull\quad
``The multi-wavelength view of hot, massive stars''; 
39$^{\rm th}$ Li\`ege Int.\ Astroph.\ Coll., 12-16 July 2010 \quad\bull}%
\def\@evenhead{}\let\@evenfoot\@oddfoot}
\makeatother
%
%
\def\beginrefer{\section*{References}%
\begin{quotation}\mbox{}\par}
\def\refer#1\par{{\setlength{\parindent}{-\leftmargin}\indent#1\par}}
\def\endrefer{\end{quotation}}
%
%
{\noindent\small{\bf Abstract:} 
Some of the Wolf-Rayet (WR) stars are found to have very high
bolometric luminosities ($\log L/L_\odot > 6$). We employ the Potsdam
Wolf-Rayet (PoWR) model atmospheres for their spectral analysis, which yields
the bolometric corrections. Distance and interstellar reddening also enter the
luminosity estimates.

Among the Galactic stars, there is a group of very luminous WNL stars
(i.e.\ WR stars of late subtype from nitrogen sequence  with hydrogen
being depleted in their atmospheres, but not absent). Their distances are 
often the major source of uncertainty. From K-band spectroscopy we found
a very luminous star ($\log L/L_\odot = 6.5$) in the Galactic center
region, which we termed the Peony Star because of the form of its
surrounding dusty nebula. A similar group of very luminous WNL stars is
found in the Large Magellanic Cloud (LMC). In the Small Magellanic Cloud
(SMC) the majority of WR stars resides in binary systems. The
single WNL stars in the SMC are not very luminous.

We conclude that a significant number of very luminous WNL stars exist
in the Galaxy and the LMC. With initial masses above 60
$M_\odot$, they apparently evolved directly to the WNL stage without
a prior excursion to the red side of the HRD. At the low metallicity of
the SMC, the binary channel may be dominant for the formation of WR
stars.}

\section{Introduction}

Very luminous stars, with bolometric luminosities exceeding $10^6
L_\odot$, appear spectroscopically mainly as Wolf-Rayet (WR) types.
However, hot stars emit most of their radiation in the extreme
ultraviolet which is not accessible to observation. Hence the
determination of their luminosity $L$ must rely on adequate model
atmospheres. Our ``Potsdam Wolf-Rayet'' code (PoWR -- see Hamann \&
Gr\"afener 2003 and references therein)   solves the non-LTE radiative
transfer in a spherically symmetric expanding atmosphere.  Detailed and
complex model atoms are taken into account especially for H, He, and the
CNO elements, while the iron-group elements are treated in the
superlevel approximation.  Wind inhomogeneities are accounted for in a
first-order approximation (``microclumping''). The code has been
applied mainly for the wind-dominated emission-line spectra of WR stars
(see  {\tt http://www.astro.physik.uni-potsdam.de/PoWR.html} for grids
of models), but can also be used for fitting photospheric absorption
spectra. In the standard version of the PoWR code, mass-loss rate and
velocity field are free parameters of the model, while they are
determined consistently with the radiation pressure only in the hydrodynamical
version (Gr\"afener \& Hamann 2005, 2008). Another PoWR code option not 
used here is ``macroclumping'' (Oskinova et al.\ 2007).

\section{How reliable are spectroscopic luminosities\,?}

\begin{figure}[b]
\centering
\epsfxsize=\textwidth
\epsffile{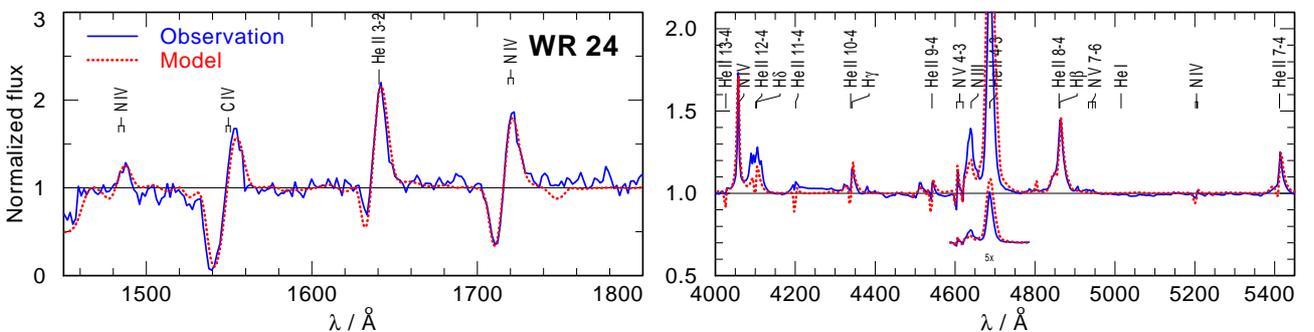}
\caption{Example of line fit for the Galactic WN6h star WR\,24, 
showing P\,Cygni type line profiles in the UV ({\em left panel}) and strong 
emission lines in the optical ({\em right}). Main parameters of the 
model (red-dotted) are $T_\ast$ = 50\,kK, $\log R_{\rm t}$ = 1.35, 
$X_{\rm H}$ = 0.44 and $v_\infty$ = 2160\,km/s. After Hamann et al.\ 
(2006)}
\label{fig:wr24-linefit}
\end{figure}

In order to discuss the reliability of WR luminosities, we briefly
describe the procedure of their spectroscopic determination. The first
step is the fit of the normalized line spectrum. Spectra from
stellar-wind models depend mainly on two parameters, the stellar
temperature $T_\ast$ and the so-called ``transformed radius'' \(R_{\rm
t}\). The terminal wind velocity $v_\infty$ controls the widths of the
profiles. Furthermore, the lines depend of course on the chemical
abundance of their species.

The stellar temperature $T_\ast$ is the effective temperature related to
the luminosity $L$ and the stellar radius $R_\ast$ via the
Stefan-Boltzmann law. $R_\ast$ refers by definition to the point of the
atmosphere where the Rosseland mean optical depth reaches 20.

The ``transformed radius'' is defined as 
\(
R_{\rm t} = R_*
  \left[\frac{v_\infty}{2500 \, {\rm km}\,{\rm s^{-1}}} \left/
  \frac{\dot M \sqrt{D}}
       {10^{-4} \, M_\odot \, {\rm yr^{-1}}}\right]^{2/3} \right. .
\)
Its name, historically coined by Schmutz et al.\ (1989), is actually
misleading since $R_{\rm t}$ has not the meaning (although the units)
of a radius. More suggestive is to consider $R_{\rm t}^{-3}$, which
might be called a ``normalized emission measure''. Being proportional
to the volume integral of the density squared, divided by the stellar
surface, $R_{\rm t}^{-3}$ scales with the emission from recombination
lines normalized to the continuum. This explains why different
combinations of $R_\ast$, $v_\infty$, and mass-loss rate $\dot{M}$
result in approximately the same normalized WR emission-line strengths
as long as $R_{\rm t}$ (or $R_{\rm t}^{-3}$) is kept at the same
value. ($D$ is the clumping factor for which we assume $D=4$
throughout the work described here.)

Taking advantage of this approximate parameter degeneracy, the analysis
can start from models with an arbitrarily adopted luminosity (our grids
are mostly calculated for $\log(L/L_\odot) = 5.3$) and find the optimum fit of
the normalized line spectrum by varying $T_\ast$ and $R_{\rm t}$ (using
models of adequate $v_\infty$ and chemical composition). The spectral
fits that can be achieved are satisfactory (cf.\
Fig.\,\ref{fig:wr24-linefit}), but also often leave characteristic
discrepancies which we attribute mainly to deviations from wind symmetry and 
homogeneity. The fit
parameters can be typically determined to an accuracy of $\pm0.05$ in
$\log T_\ast$ and $\pm0.1$ in $\log R_{\rm t}$.

\begin{figure}[t]
\centering
\epsfxsize=0.8\textwidth
\epsffile{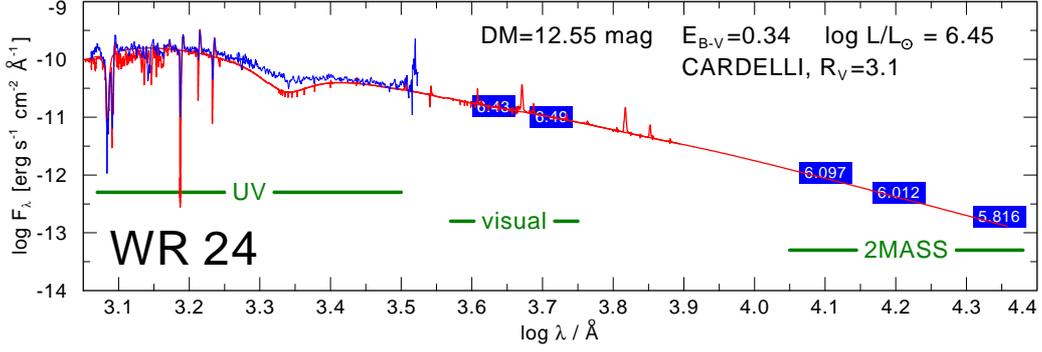}
\caption{Example of a fit of the spectral energy distribution (SED) 
for WR\,24 (red: model, blue: observation), using the model selected 
from the line fit (cf.\ Fig.\,\ref{fig:wr24-linefit}). The distance 
modulus $D\!M$ is known from cluster membership. Free parameters are the 
reddening and the luminosity scaling.
\label{fig:wr24-sed}
}
\end{figure}

In a second step, the luminosity is determined from fitting the 
spectral energy distribution of the model observations (flux-calibrated 
spectra or photometry) over the widest 
available range (see Fig.\,\ref{fig:wr24-sed}). The slope and form of 
the model SED is fitted by adjusting the color excess $E_{B-V}$, and by 
chosing an adequate reddening law (and its parameters). Sometimes the 
(circumstellar?) reddening is clearly anomalous, as demonstrated in the 
example of Fig.\,\ref{fig:wr24-sed} by the  
weak 2200\,\AA\ feature. The absolute value is adjusted by scaling the 
model in luminosity (i.e.\ a vertical shift in the double logarithmic 
plot).

In order to discuss the error margins of the obtained luminosity, 
we consider $M_{\rm bol} = 4.72 - 2.5 \log L/L_\odot$.  This 
absolute bolometric magnitude follows from the observed apparent 
magnitude $m_i$ in some band $i$ (where $i$ may stand for, e.g., 
the visual band $V$, 
or for the near-IR $K$ band in case of visually obscured objects), 
by applying the bolometric 
correction $BC_i$ and the extinction $A_i$ in the considered band, 
and the distance modulus $D\!M$ 
(all in mag),   
\begin{equation}
M_{\rm bol} = m_i - BC_i - A_i - D\!M
\end{equation}

\begin{figure}[!b]
\centering
\begin{minipage}{8.2cm}
\epsfxsize=8.2cm
\epsffile{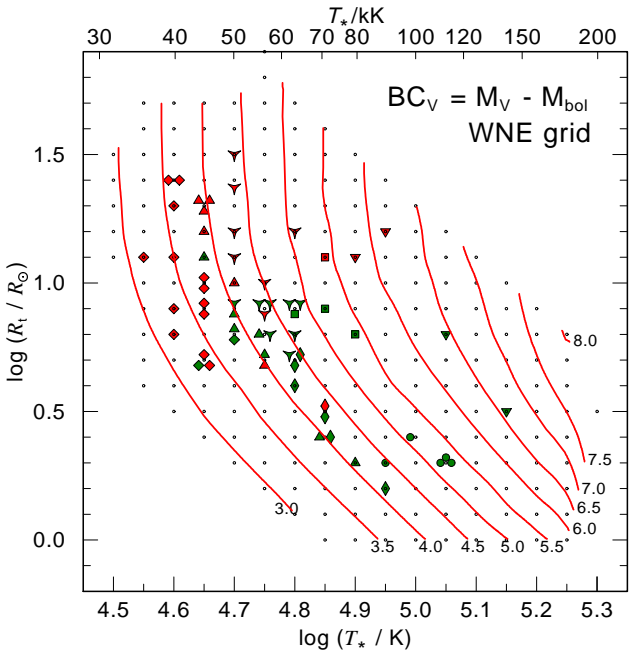}
\end{minipage}
\hspace{0.4cm}
\begin{minipage}{8.2cm}
\caption{Contours of the same Bolometric Correction to the visual magnitude 
(labels: $BC_V$ in mag)
for PoWR models of hydrogen-free WN stars.  
Small dots mark the WNE grid models,
and colored symbols show the parameters of Galactic WN stars from Hamann 
et al.\ (2006).}
\label{fig:BC}
\end{minipage}
\end{figure}

The bolometric correction is predicted by the model (see
Fig.\,\ref{fig:BC}) and amounts to 3--5\,mag for most WN stars, but can
reach 6.5\,mag for the hottest subtypes. Since the observed band lies in
the Rayleigh-Jeans domain of the SED, the bolometric correction
($10^{0.4BC}$) scales roughly with $T^3_\ast$. Hence a typical fit
uncertainty of $\pm0.05$ in $\log T_\ast$ propagates to $\pm0.4$\,mag in
$BC$ or $\pm0.15$ in $\log L$.

The extinction can introduce a noticeable error especially when it is high. If
the available wavelength basis is as long as in the example shown in
Fig.\,\ref{fig:wr24-sed}, the color excess $E_{B-V}$ can be determined from
the spectral slope to a few hundreds of a magnitude. However, the extinction
law itself can vary between different lines of sight. For the visual band, for
instance, $R_V$ may deviate considerably from 3.1 as the Galactic standard
value, which affects the visual extinction $A_V = R_V\ E_{B-V}$.

For Galactic WR stars, the largest uncertainty often comes from the 
distance. Only when a star can be assigned to an open cluster or 
association, we can adopt the distance of the latter. Stars in the 
Galactic Center, and especially objects in the LMC and SMC, have the 
clear advantage of a well-known distance. 

Summarizing, the reliability of spectroscopically determined 
luminosities of WN stars depends very much on the individual 
circumstances. When the stellar temperature is well constrained from the 
line fit of different ions, the distance is known, and good photometry 
is available in spectral bands with not too high extinction, the error 
margins combine to $\pm0.2$ in $\log L$, typically. 

\begin{figure}[!b]
\centering
\epsfxsize=\textwidth
\epsffile{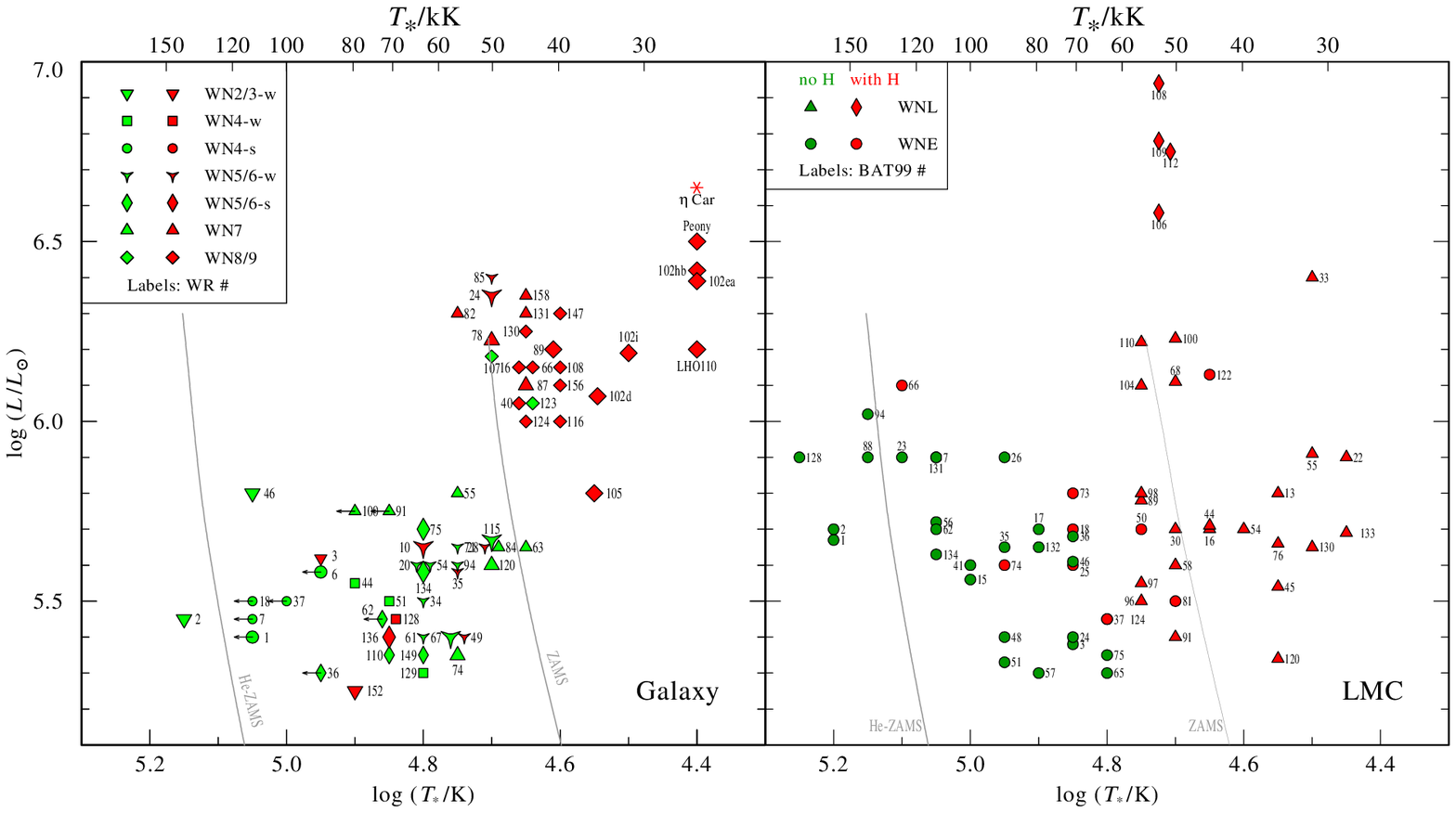}
\caption{Empirical HRDs of WN stars in the Galaxy ({\em left}) and the 
LMC ({\em right}). The Galactic stars  are from Hamann et al.\ 
(2006). Labels denote the WR catalog number, while red and green colors 
indicate whether photospheric hydrogen is detectable or not, 
respectively. The larger symbols refer to stars with independently known 
distances. Included are also the Galactic center stars WR\,102hb, 102ea,
102i, 102d and LHO\,110 analyzed by Liermann et al.\ (2010, and these
proceedings) and the Peony star (WR102ka) from Barniske et al.\ (2008).
$\eta$\,Car (after Figer et al.\ 1998) is shown for comparison. The LMC
stars in the right panel (identified by their BAT99 catalog number) are
preliminary results from R\"uhling et al.\ (in prep.). The diamonds
indicate four objects from the R136 cluster according to Crowther et
al.\ (2010), see also Schnurr et al.\ (these proceedings).}
\label{fig:HRDs}
\end{figure}

\section{Results}

The empirical Hertzsprung-Russell-diagram (HRD) for the Galactic WN
stars, analyzed with the methods described above, is
displayed in Fig.\,\ref{fig:HRDs} (left panel). Two groups of stars
are clearly distinguished. The hydrogen-free stars, usually termed WNE
(``E'' for early subtypes), are pretty hot and located between the
hydrogen and the helium main sequence. Their luminosities $\log
L/L_\odot$ are between 5.3 and 5.8, typically. In contrast, the WNL
(``L'' for late subtypes) stars are less hot than the ZAMS and contain
hydrogen. Most of them are very luminous ($\log L/L_\odot > 6$). They are 
not WR stars in the classical understanding, but rather very
massive stars with strong winds which are still in the
hydrogen-burning phase. However, for a couple of the WNL stars in this
diagram the distance is actually not known (indicated by their smaller
symbols). Their luminosity is basically adopted from similar stars of
known distance. A couple of them might be in fact closer and less
luminous.

\begin{figure}
\begin{minipage}{9.2cm}
\epsfxsize=9.2cm
\epsffile{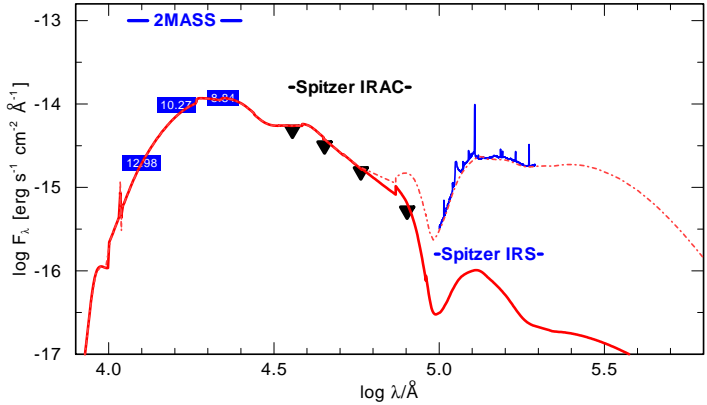}
\end{minipage}
\hspace{0.4cm}
\begin{minipage}{7.2cm}
\caption{SED fit of the Peony star (WR\,102ka) in the Galactic center
region. Because of the strong interstellar extinction, the maximum of
the observed flux is in the $K$ band. The mid-IR excess observed with {\em
Spitzer} indicates warm circumstellar dust (model: red dashed line), 
which was found here for the first time around a WN star. From Barniske
et al.\ (2008)}
\label{fig:wr102ka-sed}
\end{minipage}
\end{figure}

Remarkable is the group of WNL stars from the Galactic center region. 
Being visually obscured, they have been analyzed only from their $K$-band
spectra. Nevertheless, we believe that their luminosities are  reliable.
Figure\,\ref{fig:wr102ka-sed} shows the SED  fit for the Peony star. In
the $K$ band, the extinction $A_K$ amounts only to about 3\,mag. The analysis
(Barniske et al.\ 2008) revealed that this star has the second-highest
luminosity ($\log L/L_\odot$ = 6.5 $\pm$0.2) known in the Galaxy, after
$\eta$\,Car. 

In the LMC the group of very luminous WNL stars is less pronounced 
(Fig.\,\ref{fig:HRDs}, right panel). Instead there are many WNL stars with hydrogen,
but moderate luminosities. It is not yet clear if we have missed such
stars in the Galaxy because of the distance problem discussed above. The
four extremely luminous stars represented 
in the figure by diamonds have been claimed recently by
Crowther et al.\ (2010), and 
belong to the cluster R136 which we had avoided. The
problem is to rule out accidental multiplicity in this very dense
field of stars with identical spectral type.  

The 12 WR stars (11 WN, 1 WO) known in the SMC are currently studied 
by Pasemann et al.\ (these proceedings, and in prep.). 
5 of these stars appear to be single, while the remaining 6 are binaries
for which we started to analyze the composite spectra. 
With $\log L/L_\odot < 6$ the single WN stars are not ``very'' luminous.

\section{Conclusions}

A couple of very luminous stars ($\log L/L_\odot > 6$) are found in the
Galaxy and LMC. Apart from a few outstandingly bright LBVs, the 
most-luminous stars are of late WN
type (WNL). Remarkable is the abundance of such stars in the Galactic
center region. In the SMC, single WNL stars are also found despite of the
low metallicity, but not with very high luminosities.

%
%
%
%
\footnotesize
\beginrefer

\refer Barniske A.,  Oskinova L.M., Hamann W.-R., 2008, A\&A, 486, 971

\refer Crowther P.A., Schnurr O., Hirschi R., et al., 2010, MNRAS (in press)

\refer Figer D.F., Najarro F., Morris M., et al., 1998, ApJ, 506, 384

\refer Gr\"afener G., Hamann W.-R., 2005, A\&A, 432, 633

\refer Gr\"afener G., Hamann W.-R., 2008, A\&A, 482, 945 

\refer Hamann W.-R., Gr\"afener G., 2003, A\&A, 410, 993

\refer Hamann W.-R., Gr\"afener G., Liermann A., 2006, A\&A, 457, 1015

\refer Liermann A., Hamann W.-R., Oskinova L.M., Todt H., Butler K., 
       2010, A\&A, in press

\refer Oskinova L. M., Hamann W.-R., Feldmeier A., 2007, A\&A, 476, 1331

\refer Schmutz W., Hamann W.-R., Wessolowski U., 1989, A\&A, 210, 236 

\endrefer

\end{document}